\documentclass[twocolumn,aps,floats,floatfix,superscriptaddress]{revtex4}
\usepackage{graphicx,amssymb}
\usepackage{epsfig}
\newcommand{\beq}{\begin{equation}}
\newcommand{\eeq}{\end{equation}}

\newcommand{\bE}{{\bf E}}

\newcommand{\bC}{{\bf C}}

\newcommand{\bX}{{\bf X}}

\newcommand{\bZ}{{\bf Z}}

\newcommand{\bL}{{\bf L}}
\newcommand{\br}{{\bf r}}
\newcommand{\bj}{{\bf j}}

\begin{document}
\draft
\title{Hidden information in fluctuation in small systems}
\author{ Ruijie Qian}
\affiliation{Surface Physics Laboratory, Department of Physics, Fudan University,
Shanghai 200433, China\\
}
\affiliation{Institute of Technical Physics, Shanghai, China \\
}
\author{S. T. Chui}
\email{chui@udel.edu}
\affiliation{Bartol Research Institute and Dept of Physics and Astronomy, University of Delaware, Newark, DE 19716, USA\\
}
\author{Zhenghua An$^1$}
\email{anzhenghua@fudan.edu.cn}
\author{Hongtao Xu}
\affiliation{Fudan University Nanofabrication Lab\\}
\author{Zhifang Lin}
\author{Zian Ji}
\affiliation{Surface Physics Laboratory, Department of Physics, Fudan University,
Shanghai 200433, China\\
}
\affiliation{Key Laboratory of Micro and Nano Photonic structures (Ministry of
Education), Fudan University, Shanghai, China\\
}
\author{Wei Lu}
\email{luwei@mail.sitp.ac.cn}
\affiliation{Institute of Technical Physics, Shanghai, China \\
}
\affiliation{ Shanghai Tech, China \\
}
\begin{abstract}
The exploration of the rich dynamics of electrons is a frontier in fundamental nano-physics. The dynamical behavior of  electrons is dominated by random and chaotic thermal motion with ultrafast ($\approx$ ps) and nanoscale scatterings. This generates  fluctuating electromagnetic fields in close vicinity of the moving electrons. W studied this fluctuation in small structures and found that its spatial distribution is not uniform, the magnitude of the fluctuation depends on external parameters such as the size ( ~1 $\mu m$) and the shape of the structure and changes can occur by an order of magnitude. Our work opens the possibility of improving the signal to noise ratio in small devices and in manipulating microscopic electron kinematics through nano-optical techniques and to applications in thermal detectors and photothermal photovoltaics. 
\end{abstract}
\maketitle
In sensors and circuits, the charge fluctuation determines the sensitivity and effectiveness of the device\cite{19}. 
The near field transfer of energy that exceeded the black body limit\cite{1,2,3,4,5}  and the Casimir force\cite{6,7,8,9,10,11}  are determined by the fluctuation of the electromagnetic (EM) field \cite{12,13,14,15,16}  which in turn is determined by the charge/polarization density fluctuation of the objects involved\cite{12,17}. Radiative transfer of energy at a small length scale is of great importance to different technologies including heat assisted magnetic recording\cite{18}, near-field thermophotovoltaics\cite{14} and lithography\cite{15}.  The emergence of nanotechnology leads naturally to questions about the fluctuations in small systems. While some studies have been carried out to understand the fluctuation in large structures\cite{20,21} no work has been done to study in small structures the spatial and frequency distribution of the fluctuation which are affected by external conditions such as the size and the shape of the system. Related researches on shot noise in two dimensional electron gas (2DEG) and Johnson-Nyquist noise in quantum-voltage-noise source Johnson noise thermometer(QVNS-JNT) system have been reported\cite{22}, suggesting that noise is of great use in accurate measurements. However, the more usual circumstances of fluctuation under thermal equilibrium have been ignored due to the difficulties in its measurement. As a first step, here we reported the study of the spatial charge fluctuation  at a fixed frequency for micron size gold disks and triangles theoretically and experimentally. The average amplitude of the charge fluctuation is zero. Thus our work differs from conventional studies of the response from external fields. We found that the mean square charge fluctuation in a small metallic disk can be cylindrically symmetric.  A cylindrically symmetric charge distribution does not emit (far field) radiation. They cannot be observed with a lens and can only be observed in the near field. Even then the signal is very weak and requires a very special sensitive sensor which enables us to observe a nontrivial spatial dependence of the fluctuation for the first time. This cylindrical charge fluctuation cannot be generated by external electric fields in the same frequency range. These features distinguish it from some recent experimental results\cite{oe1}. We found that the magnitude of the fluctuation can change by an order of magnitude when the size of the disk is changed. Agreement is found between theory and experiment.
 Our work opens the door to the design, to manipulating the fluctuation for applications and to improving the sensitivity of small devices.

We first describe our theoretical result for the charge fluctuation distribution.
In metals the dominant noise is the Johnson noise which can be interpreted as due to locally fluctuating electric fields\cite{12} $E^n(\br,\omega)$ of angular frequency $\omega$ at position $\br$
so that their averages and mean sqaure averages are given by $<\bE^{n}(\br,\omega)>=0$; 
\beq 
<\bE^{n *}(\br,\omega)\cdot \bE^n(\br'\omega')>=e\delta(\omega-\omega')\delta (\br-\br')
\label{core}
\eeq
for a constant $e$ determined by the sum of the quantum and the thermal fluctuation that is proportional to  the temperature.  
In a finite metallic film the fluctuating fields induce fluctuating surface charge densities which interact with each other at {\bf finite frequencies} as is described by Maxwell's equations and not just by Coulomb's law. 
We have recently developed a formulation\cite{book,cone,disk,tri,detail0} of solving Maxwell's equations 
that can be faster than existing approach by three orders of magnitude \cite{ms}. This involves a new way to implement the boundary condition at the edge; the physical quantities are expressed not on a mesh but in terms of a complete orthonormal set of basis functions functions $\bX_{j}(r)$ labelled by the index j. 
For a disk, $\bX$ is the vector cylindrical basis function that is proportional to the function $\exp(im\phi) J_m(x)$ of angular momentum m\cite{disk,detail0}.
$J_m$ is the Bessel function, $x=k_{j}r$; the discrete set of wave vectors $k_j$ are picked so that at the boundary $r=R$, the radial derivative is equal to zero. 
The noise electric field can be expanded in the basis as
$\bE^n(r)=\sum E^n_{X,j}\bX_{j}(r).$ The current density can be similarly expanded. 
By Ohm's law, in terms of the resistivity $\rho$ and the current density $\bj$, $\rho\bj=\bE_{tot}$ where the {\bf total  local} electric field is a sum of the external field $\bE_{ext}$, the electric field from the finite frequency electron-electron interaction due to the currents at other places
$\bE_{em}$ and a boundary field $\bE_s$ due to charge accumulated at the boundary of the finite sample\cite{book,disk,tri,cone,detail0}. 
The finite frequency electron-electron interaction from Maxwell's equations in integral form can be written as $\bE_{em}=-\bZ_0\bj$  where the impedance matrix $\bZ_0=i\omega\mu_0(\bL-\omega^{-2}/\bC) $  comes from the inductance $\bL$ and the capacitance $\bC$ which are just representations
of the Green's function in the orthonormal basis\cite{book,disk,tri,cone,detail0}. We thus obtain:
$\bZ\bj=\bE_{ext}+\bE_{s}.$
where $\bZ=\bZ_0+\rho$, $\bE_s$ is a field localized at the boundary. It is determined by the condition that the component of the current perpendicular to the boundary is zero. 

For the present case, the external field is from the fluctuation field with zero averages:  
$j_{i}= \sum_j Z^{-1}_{ij} (E^n_j+E_{sj}).$
From Eq. (\ref{core}) and the charge current conservation,
the mean square average of the expansion coefficient for the current and the charge density inside the sample at angular frequency $\omega$ is thus given by\cite{detail0} 
$\omega^2<n_{ai}n_{bj}^*>=<k_ik_jj_{ai}j^*_{bj}>=\sum_i k_ik_jZ_{ai}^{-1} Z_{bi}^{* -1} e^2.$

For a disk of radius R, because of the cylindrical symmetry, the impedance matrix is diagonal in m. 
We study samples with thicknesses $t$ less than the skin depth. The two dimensional charge density fluctuation $S^2(r)=<n^2(r)>$ is proportional to the thickness.  It comes from contributions of different 
angular momenta m. For a given m, they are from contributions of different wave vectors:\cite{detail0}:
\beq
S^2(r)\propto e^2\sum_{m,k,p} O_m(k,p,r)/\omega^2
\label{finalS}
\eeq 
$O_m(k,p,r)=\sum_{k'} Z_{m,k,k'}^{-1} Z_{m,p,k'}^{-1}  kp J_{m}(kr)J_{m}(pr)$
is a "susceptibility" in terms of the Bessel function $J_m$; for different m it contributes to the fluctuation at different distances away from the center. Only the m=0 component contributes to the fluctuation at the center $r=0$.
To explore which m provides for the interesting feature, we have kept
the maximum angular momentum channel as $m_{max}=15$. For each m, we find that
8 wave vectors for each m is  adequate\cite{disk}.

\begin{figure}[tbph]
\vspace*{0pt} \centerline{\includegraphics[angle=0,width=5cm]{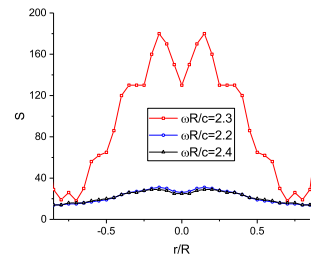}} 
\caption{Root mean squared charge density fluctuation S of a disk 
of radius R at different dimensionless frequencies for a resistivity $\rho=0.01$ in units of $\rho_u$ .}
\label{resr}
\end{figure}
An interesting behaviour is manifested in the frequency dependence of the spatial fluctuation distribution when the radius is comparable to the wavelength. Fig. (\ref{resr})
shows the theoretical signal $S(r)$  in arbitrary units across a diagonal as a function of the normalized radial distance $r/R$ at different angular frequencies 
for a finite resistivity  $\rho=0.01\rho_u$ ( $\rho_u=Z_0\lambda /tR$ where $Z_0$ is the resistance of the vaccum, 277 $\Omega$, $\lambda$ is the wavelength.)
When $\omega R/c=2.3$ there is a big increase in the fluctuation strength.  The signal exhibits a dip in the center. The smaller the resistivity, the bigger the peak. The jump appears because of the frequency dependent balance between the capacitive and the inductive effect so that the effective response, which is measured by the inverse of the smallest eigenvalue of the impedance matrix, becomes large at some frequency. A similar behaviour is also observed experimentally. We describe this next.

\begin{figure}[tbph]
\vspace*{0pt} \centerline{\includegraphics[angle=0,width=7cm]{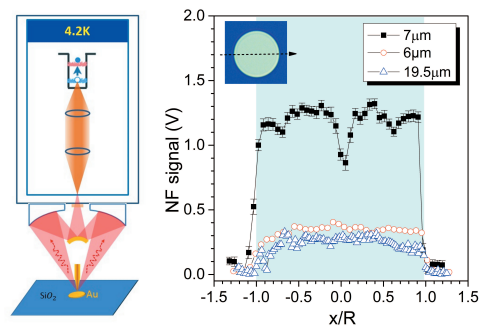}} 
\vspace*{0pt}
\caption{Left: The experimental setup of our near field noise microscope. Light emitted by the AFM tip and modulated by a tapping frequency is captured by a confocal system and detected by sensors kept at 4.2K. Right:The experimental signal along  diameters of gold disks of different sizes.}
\label{setup}
\end{figure}
To map the charge fluctuation, we use the scanning noise microscope (SNoiM)\cite{22,A}, which is a modification of the scattering scanning near-field optical microscope (s-SNOM)\cite{B}.
For the traditional s-SNOM, the scattered light from an atomic force microscope tip is analyzed. External illumination is essential. 
To detect a signal induced only by the charge fluctuation prohibits us from applying an external illumination. In our work, light representing the fluctuating EM evanescent wave are scattered by  tungsten atomic force microscope (AFM) tips with diameters 50 to 100nm, then collected by a confocal optical system and detected by a charge-sensitive infrared phototransistor (CSIP).  The height of tungsten tip is modulated by a piezoelectric ceramics with a frequency of 5 Hz; 
the near-field signal is demodulated at 5Hz with a lock-in amplifier. The shear force between the tungsten tip and metal surface is controlled by a feedback loop to avoid the contact between tip and sample. The excellent performance of CSIP (about 2 orders of magnitude more sensitive than traditional HgCdTe detectors) together with the demodulation at the tapping frequency make it possible for us to extract the extremely faint near field signal from the much larger far-field signal.
Our system is illustrated in Fig.\ref{setup}. It has been  mentioned in our previous work\cite{22}. We next explain how the experimental signal is related to the charge fluctuation. 

We have recently analyzed the s-SNOM signal by studying the scattering of the electromagnetic (EM) waves from the conical shape AFM tip. The singularity of the cone tip is taken care of with basis functions which contains this integrable singularity for the cone obtained from the
conformal mapping between the cone and a circle\cite{cone}. Whereas previous calculations can obtain agreement with experiment only to within thirty per cent, our result achieve agreement to within a few per cent. Our physical picture is that the external scattering field induces an oscillating charge at the tip, which then interacts with the sample in near field via the Coulomb interaction. With this we obtain good agreement between theory and experiment for the charge density distribution induced on a gold disk by an external EM field\cite{jpcm}.
This result is directly applicable to SNoiM.
The  fluctuating surface charge density on the disk interacts via the Coulomb interaction with the AFM tip which give the SNOM signal\cite{noism}. 
The mean square fluctuation of the SNoiM signal 
is thus proportional to mean square charge fluctuation $<S^2(\br)>$. The SNoiM signal in principle also contains a contribution from the magnetic fluctuation on the disk. We find that\cite{detail0} this contribution is usually smaller by more than three orders of magnitude. 

The SNoiM signal as a function of position across disks of different diameters on samples fabricated by electron beam lithography and  thermal evaporation of 5nm Cr and 100nm Au on substrates of silicon dioxide and silicon is shown in Fig. \ref{setup} with the detector tuned to focus on radiation with a fixed wavelength of 14.1 $\mu m$. There is a sudden increase in signal strength when the diameter of the disk is 7 $\mu m$. A dip is exhibited in the signal at the center. These features are in agreement with the theoretical results in Fig. 1. The noise is the same for all the experimental curves in Fig. 2.  It is not possible to see any structures when the signal is small and comparable to the noise. We believe that the difference in the shape between the theory and the experiment comes from the non-uniformity of the experimental sample.

\begin{figure}[tbph]
\vspace*{0pt} \centerline{\includegraphics[angle=0,width=7cm]{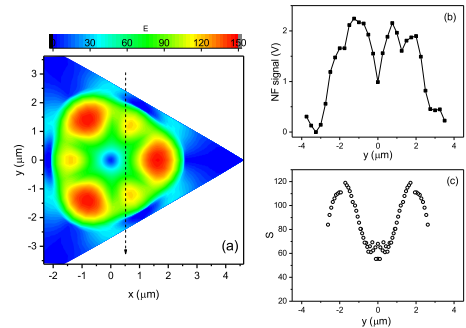}}
\vspace*{0pt}
\caption{(a)Contour plot of the theoretical root mean square charge fluctuation in an equilateral triangle in arbitrary units at a dimensionless angular frequency $\omega a/c=3.4$ (b) The experimental signal as a function of y for a triangle of base length a=8 $\mu m$ at a wavelength 14.1 $\mu m$ scanned along the line indicated in (a). (c) Theoretical result as a function of y at x=0.125$\mu m$ and x=0.075$\mu m$}
\label{trie}
\end{figure}

To confirm the generality of the nonuniform distribution of the fluctuation, we have also performed exploration of the fluctuation distribution in finite triangular films. Our experimental result at a wavelength of 14.1 $\mu m$ is shown in Fig. (\ref{trie})(b). This result is for a scan 
along the line indicated in (a). The base of the triangle is of length 8$\mu m$. We discuss next the comparison with theoretical results.    

Our calculation uses a set of orthonormal basis functions. We have recently studied the scattering of EM waves from triangular films with basis functions obtained from a conformal mapping between the triangle and the circle\cite{tri}. With this approach, we have computed the spatial distribution of the mean square charge fluctuation in an equilateral triangle of sides of length $a$. Our theoretical result for the two dimensional distribution of this fluctuation at an angular frequency $\omega=3.4 c/a$ is shown in Fig. \ref{trie}(a). Scans of the signal similar to that in the experimental measurement is also shown in (c). This is similar to that observed experimentally.
\begin{figure}[tbph]
\vspace*{0pt} \centerline{\includegraphics[angle=0,width=5cm]{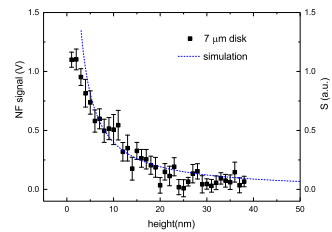}} 
\vspace*{0pt}
\caption{ Symbol: Experimental  signal as a function of the tip height. Line: The root mean squared of the component of the Coulomb electric field perpendicular to the sample averaged over the AFM tip as a function of the distance from the sample.}
\label{hd}
\end{figure}

The near field signal decreases as the tip is moved away from the sample. The experimental dependence of the signal on the tip height is shown in Fig. \ref{hd}. As we explained above, in our picture this depends on the magnitude of the average electric field due to the samples charges at the tip. The near electric field is just the Coulomb field. The theoretical result obtained from the functional dependence of root mean squared of the component of the Coulomb field\cite{inti} perpendicular to and averaged over  the sample is also shown in Fig. \ref{hd}.
Reasonable agreement is found between the theoretical and the experimental results.

\begin{figure}[tbph]
\vspace*{0pt} \centerline{\includegraphics[angle=0,width=7cm]{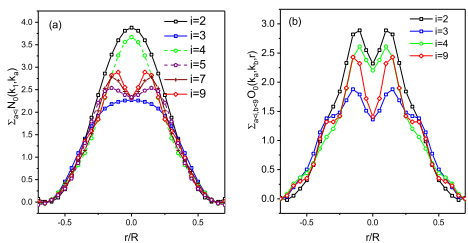}} 
\caption{ Left: Contribution to  the "susceptibility" $\sum_{a<i} O_0(k_a,k_1)$
at different positions from  m=0 but different i. Right: Contribution to  $\sum_{a<i,b<9} O_0(k_a,k_b,r)$
at different positions from  m=0 but different i. }
\label{diskm0f}
\end{figure}

The signal for the disk exhibits a dip at r=0 which we now explain. The fluctuation is given by Eq.(\ref{finalS}). 
The signal is dominated by contributions around the origin $r=0$ , which can only come about from the m=0 terms. This indicates that around the center Eq(\ref{finalS}) is dominated by the contribution with $m=0$.  
We show in the left panel of Fig. \ref{diskm0f} the partial sum 
$\sum_{a<i} O_0(k_a,k_1,r)$ at different positions from the m=0 contribution when the sum over one of the wave vectors is taken up to less than $k_i$ while the other k is fixed at its lowest value.  
As can be seen there is no dip for small i. The dip appears when more rapidly varying contributions with larger k are included. Physically, charge fluctuation that are radially more rapidly varying (larger k) provides for larger capacitive terms which makes the impedance matrix elements for these contributions negative. These negative contributions dies off rapidly with distance and thus only lowers the response close to the center 
This dip is maintained when more values of k are included, as is shown in the right panel of  Fig. \ref{diskm0f} where we show the partial sum $\sum_{a<i,b<9} O_0(k_a,k_b,r)$ with different wave vector combinations 
where  with one of the wave vectors kept at the maximum value and the number of the other wave vectors kept is increased. The results here with only the mode with angular momentum m=0 kept is close to that in the small r region of the final result in Fig. 1 when all values of m are included. 

\begin{figure}[tbph]
\vspace*{0pt} \centerline{\includegraphics[angle=0,width=5cm]{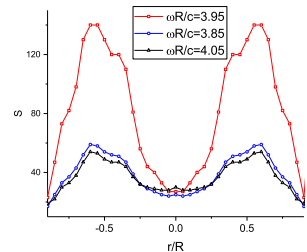}} \vspace*{0pt}
\caption{The mean square charge fluctuation S in arbitrary units for a disk at different but higher frequencies for a resistivity $\rho=0.01$ in units of $\rho_u$ .}
\label{diskhf}
\end{figure}

Similar behaviour is manifested at other frequencies.
When the frequency is increased by a factor of approximately two, 
another increase in fluctuation is manifested  for the signal for a disk but this time not 
close to the center. This is illustrated in Fig. \ref{diskhf}.
An examination of the details shows that the fluctuation is dominated by a different angular momentum channel m=1. 

In conclusion 
we find that the spatial fluctuation distributions in small systems such as metallic disks and triangles are not structureless,  because of the interaction between the charges.
As the size of the structure
is varied, the charge fluctuation can change by an order of magnitude. Agreement is found between theoretical calculations  and experimental measurements. In sensor applications, the response of systems to external signals can also become large but because the coupling and the symmetry is different, these increases do not happen under the same conditions as for the fluctuations. 
The 
signal to fluctuation ratio can be optimized.  Our work opens the door to  manipulating the fluctuation for applications, the design and the improvement of the sensitivity of small devices.

\begin{acknowledgments}
ZA was supported by National Key Research Program of China under Grant No. 2016YFA0302000, National Natural Science Foundation of China (Grant Nos. 61521005/11674070/11634012/11427807,11991060), and Shanghai Science and Technology Committee under Grant Nos. 18JC1420402 and 18JC1410300. Part of the experimental work was carried out at the Fudan Nanofabrication Laboratory. 
ZL was supported by National Key R\&D Program
of China (Grant Nos. 2018YFA0306201 and 2016YFA0301103) and
National Natural Science Foundation of China (Grant No. 11574055).
\end{acknowledgments}


\begin{thebibliography}{99}
\bibitem{noism} The current experimental setup corresponds to a zero net surface field $E_s=0$ in [24] because there are no external EM field.
\bibitem{inti}
For elements in the shape of disks
$I(x)=\int d\phi \int_0^{R_2}  r'dr'rdr E_z(r,r',2x)/(\pi R_2^2)$
$E_z(r)=x[x^2+r^2+r^{\prime 2}-2rr'\cos\phi]^{3/2}/(2\pi b^2).$
A factor of $2\pi$ from the angular integration is cancelled by a factor 
of $4\pi$ in the denominator from Gauss' law; the factor of $\pi b^2$ 
is from the averaging normalization.
\end{thebibliography}
\end{document}